\documentclass[conference]{IEEEtran}
\IEEEoverridecommandlockouts
\ifCLASSINFOpdf
\else
\fi
\usepackage[top=0.75in, bottom=0.75in, left=0.75in, right=0.75in]{geometry}
\usepackage{bbm}
\usepackage{verbatim}
\usepackage{graphicx}
\usepackage{cite}
\usepackage{url}
\usepackage[cmex10]{amsmath}
\usepackage{amssymb}
\usepackage{amsmath}
\usepackage{algorithm, algorithmicx, algpseudocode}
\usepackage[caption=false,font=footnotesize]{subfig}
\usepackage{color}
\usepackage{cite}
\usepackage{epstopdf}
\usepackage{calc}
\usepackage{array}

\newcommand{\tx}[0]{\text{Tx}}
\newcommand{\rx}[0]{\text{Rx}}
\newcommand{\hermitian}[0]{\text{H}}
\newcommand{\transpose}[0]{\text{T}}

\newcommand{\veta}[0]{\boldsymbol{\eta}}

\makeatletter
\def\ps@IEEEtitlepagestyle{%
  \def\@oddfoot{\mycopyrightnotice}%
  \def\@evenfoot{}%
}
\def\mycopyrightnotice{%
  {\footnotesize 978-1-5386-3512-4/18/\$31.00 \copyright 2018 IEEE\hfill}
  \gdef\mycopyrightnotice{}
}

\begin{document}
%
\title{Tracking Sparse mmWave Channel: Performance Analysis under Intra-Cluster Angular Spread}

\author{
\IEEEauthorblockN{Han Yan, Veljko Boljanovic, and Danijela Cabric}
\IEEEauthorblockA{Electrical Engineering Department, University of California, Los Angeles\\
Email: \{yhaddint, vboljanovic\}@ucla.edu, danijela@ee.ucla.edu}
\thanks{This work is supported by NSF under grant 1718742.}
}


\IEEEspecialpapernotice{(Invited Paper)
}

\maketitle

\begin{abstract}
Millimeter-wave (mmWave) systems require a large number of antennas at both base station and user equipment for a desirable link budget. Due to time varying channel under user mobility, up-to-date channel state information (CSI) is critical to obtain the required beamforming gain.. The mmWave sparse multipath channel is commonly exploited in designing tracking algorithms but practical angular spread is often overlooked. In this work, we study the performance bound of tracking accuracy in sparse mmWave channel that includes intra-cluster angular spreads. Power gain from angle-steering-based beamforming using tracked CSI is then analyzed. The theoretical study provides a design guideline beam-width in angle-steering under different intra-cluster angular spreads. We verify the results with two common tracking algorithms including sector beam tracking and maximum likelihood channel tracking. 
\end{abstract}


%
\IEEEpeerreviewmaketitle

\section{Introduction}
\label{sec:Introduction}
Millimeter-wave (mmWave) communications is a promising technology for future mobile networks due to abundant bandwidth. The standardization organization 3GPP has included mmWave communication in the fifth generation of mobile networks, 5G New Radio (5G-NR) \cite{5GNR_rel15}. As shown in both theory and practical testing, a mmWave system requires beamforming with large antenna arrays at both base station (BS) and user equipment (UE) to overcome severe propagation loss \cite{Rappaport:mmWavewillwork}. The directional transmission and reception require estimation and tracking of the wireless channel, so that antenna arrays can effectively provide beamforming gain. 

The mmWave channel tracking has been investigated in \cite{7390855,HKG+14_perburtation_tracking,MRM16_GD_tracking,GCY+15,PDW17}. In sector tracking, BS and UE track indices of sector beams that provide highest gain. It is used in IEEE 802.11ad \cite{6979964} and was studied by \cite{7390855} in mobile network. Works \cite{HKG+14_perburtation_tracking,MRM16_GD_tracking,GCY+15,PDW17} exploit mmWave sparse scattering, where CSI is tracked by tracking angle of arrival (AoA), angle of departure (AoD), and gain of multipath components. Such approach allows a more precise angle steering than sector tracking, and it supports advanced multiplexing precoding, e.g., eigenmode transmission. However, these prior works design and evaluate tracking algorithm in a sparse channel model formed by a superposition of a few multipath rays. In a realistic mmWave channel, multipaths exhibit clustered nature and there are non-negligible angular spreads (AS) along the dominant propagation directions \cite{6834753}. Such behavior is better modeled by a clustered sparsity model \cite{3GPP_model}.

In this work, we focus on a clustered sparse channel and provide a theoretical analysis of channel tracking accuracy and associated beamforming gain using angle steering. We derive the lower bound of channel parameter tracking accuracy under clustered multipath model, and we verify it with maximum likelihood (ML) based algorithm from literature. We then provide analysis of beamforming gain of angle steering with respect to beam-width, tracking error, and AS.  

The rest of the paper is organized as follows. In Section~\ref{sec:system_model}, we introduce the system model. Section~\ref{sec:CRLB} includes the performance analysis in terms of accuracy of propagation angle tracking and associated beamforming gain under intra-cluster angular spread. We describe two existing tracking algorithms in Section~\ref{sec:algorithm}, and numerically verify our analysis in Section~\ref{sec:simulation results}. Finally, Section~\ref{sec:Conclusion} concludes the paper.

\textit{Notations:} Scalars, vectors, and matrices are denoted by non-bold, bold lower-case, and bold upper-case letters, respectively, e.g. $h$, $\mathbf{h}$ and $\mathbf{H}$. The element in $i$-th row and $j$-th column in matrix $\mathbf{H}$ is denoted by $\{\mathbf{H}\}_{i,j}$. Transpose and Hermitian transpose are denoted by $(.)^{\transpose}$ and $(.)^{\hermitian}$, respectively. The $l_2$-norm of a vector $\mathbf{h}$ is denoted by $||\mathbf{h}||$. $\text{diag}(\mathbf{A})$ aligns diagonal elements of $\mathbf{A}$ into a vector, and $\text{diag}(\mathbf{a})$ aligns vector $\mathbf{a}$ into a diagonal matrix.

%
%
\section{System Model}
\label{sec:system_model}

\begin{figure}[t]
\subfloat[Illustration of mmWave MIMO channel with $L=1$ multipath cluster.]{%
  \includegraphics[clip,width=1\columnwidth]{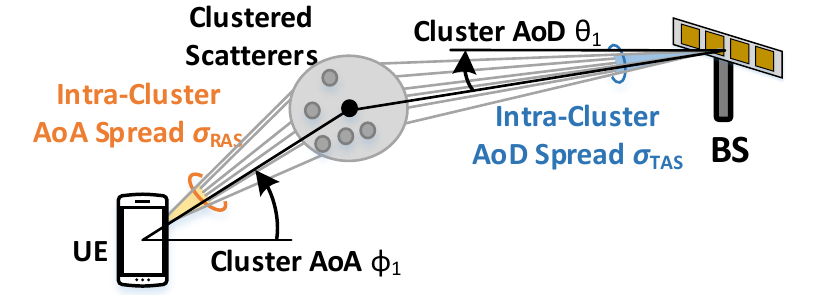}%
}
\vspace{-0mm}
\subfloat[The time frame with DL control phase (grey) and DL data phase (blue).]{%
  \includegraphics[clip,width=1\columnwidth]{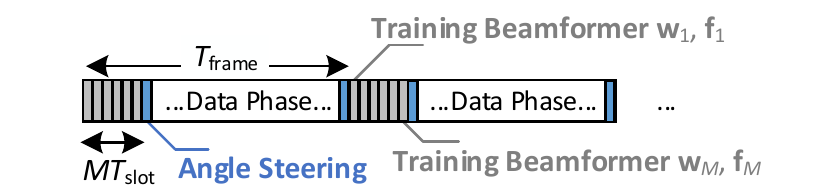}%
}
\caption{System model of tracking based mmWave system.}
\vspace{-3mm}
\label{fig:system_model}
\end{figure}

We consider a system in the downlink (DL) with a BS transmitter with $N_{\tx}$ antenna and a UE receiver with $N_{\rx}$ antennas. The multiple-input and multiple-output (MIMO) channel consists of $L$ multipath clusters, and each of them have $R$ intra-cluster rays \cite{3GPP_model} as shown in Fig.~\ref{fig:system_model}(a). We focus on narrowband model in the azimuth plane, and the channel matrix $\mathbf{H}$ is expressed as

\begin{align}
\mathbf{H} = \frac{1}{\sqrt{LR}}\sum_{l=1}^{L}\sum_{r=1}^{R}g_l e^{j\psi_{l,r}}\mathbf{a}_{\text{Rx}}(\phi_l+\Delta\phi_{l,r})\mathbf{a}_{\text{Tx}}^{\text{H}}(\theta_l+\Delta\theta_{l,r})
\label{eq:channel_model}
\end{align}
In the above equation, $\mathbf{a}_{\rx}(\phi)\in \mathbb{C}^{N_{\rx}}$ and $\mathbf{a}_{\tx}(\theta)\in \mathbb{C}^{N_{\tx}}$ are spatial responses corresponding to AoA $\phi$ and AoD $\theta$. In uniform linear array (ULA) with half-wavelength antenna spacing, their $k^{\text{th}}$ elements are $\{\mathbf{a}_{\tx}(\theta)\}_k = e^{j\pi(k-1)\sin(\theta)}$ and $\{\mathbf{a}_{\rx}(\phi)\}_k = e^{j\pi(k-1)\sin(\phi)}$, respectively. Cluster-specific parameters $\phi_l$, $\theta_l$, and $g_l$ correspond to the AoA, AoD, and path gain of $l^{\text{th}}$ cluster, respectively. Ray-specific parameters include intra-cluster angular offset (AO) in AoA $\Delta\phi_{l,r}$ and AoD $\Delta\theta_{l,r}$, and complex gain $e^{j\psi_{l,r}}$ of the $r^{\text{th}}$ ray in that cluster, where $\psi_{l,r}$ is uniformly distributed within $(-\pi,\pi]$. Each of the AO is an independent and identically distributed (i.i.d.) random variable with known probability density function (PDF) $p_{\text{AO}}(.)$, with zero mean and variances $\sigma^2_{\text{TAS}}$ and $\sigma^2_{\text{RAS}}$ for transmitter and receiver sides, respectively. We define $\sigma_{\text{RAS}}$ and $\sigma_{\text{TAS}}$ as angular spreads in AoA and AoD. In this work, we focus on channel with $L=1$ cluster\footnote{We consider both line-of-sight (LOS) and non-LOS (NLOS) cluster. The former has zero AS and one ray while the latter has non-zero AS. The results would apply to channels with $L>1$ clusters due to the decoupling \cite{DBLP:journals/corr/Abu-ShabanZASW17}.} and denote its cluster-specific parameters as $\phi_{\text{c}},\theta_{\text{c}},g_{\text{c}}$, and ray-specific parameters as $\Delta\phi_r, \Delta\theta_r, \psi_r$. We define vectors $\veta_{\text{c}} = [\phi_{\text{c}},\theta_{\text{c}},g_{\text{c}}]^{\transpose}$ and $\veta_{\text{r}} = [\Delta\phi_1, \Delta\theta_1, \psi_1 \cdots, \Delta\phi_R, \Delta\theta_R,\psi_R]^{\transpose}$.


We focus on two phases in DL: 1) control phase with channel tracking, and 2) data phase with beamformed data communication using acquired CSI as shown in Fig.~\ref{fig:system_model}(b). The DL control phase contains $M$ time slots every $T_{\text{frame}}$. In the $m^{\text{th}}$ slot (with fixed duration $T_{\text{slot}}$), the transmitter uses a training precoder $\mathbf{f}_m\in \mathbb{C}^{N_{\tx}}$ and the receiver\footnote{We focus on UE receiver with single RF-chain, i.e., analog beamforming receiver, as typically considered for cost concern. We also assume that analog combiner has both phase and magnitude control capability \cite{6170865}.} uses a training combiner $\mathbf{w}_m \in \mathbb{C}^{N_{\rx}}$. Both precoder and combiner have unit power, i.e., $\|\mathbf{f}_m\| = \|\mathbf{w}_m\|=1,\forall m$. Assuming perfect synchronization, unit pilot symbol and channel does not change during training, the received signal $\mathbf{y}\in \mathbb{C}^{M}$ is 
\begin{align}
&\mathbf{y} 
= \text{diag}\left(\mathbf{W}^{\text{H}}\mathbf{H}\mathbf{F}\right) +\mathbf{n},
\label{eq:received_signal_model}
\end{align}
where $\mathbf{F} = [\mathbf{f}_1,\cdots,\mathbf{f}_M]$ and $\mathbf{W} = [\mathbf{w}_1,\cdots,\mathbf{w}_M]$ are training beamformers at the BS and UE, respectively. We denote thermal noise power at each receiver antenna as $\sigma_\text{n}^2$, and the post-combining noise $\mathbf{n}\in \mathbb{C}^{M}$ is Gaussian random vector\footnote{Noise in independent in different time slots.}, i.e., $\mathbf{n} \in \mathcal{CN}(0,\sigma_\text{n}^2\mathbf{I}_M)$.

Additionally, with practical period $T_{\text{frame}}$, e.g., $\sim$10ms, we assume the ray-specific parameter $\veta_{\text{r}}$ lose time coherence, and their realizations are independent in data phase slots. However, the cluster-specific parameters $\veta_{\text{c}}$ are slow varying within $T_{\text{frame}}$, and we assume they are close to $\veta_{\text{c}}$ from the previous time frame\footnote{In LOS with 50m between BS and UE, 60mile/hr speed results in $0.6^\circ$ change. In NLOS with scatterers 5 meter from UE, 2m/s speed results in $0.4^\circ$ AoA change. Rotation of UE may result in a higher angle change.}. For mathematical tractability, we study data phase gain with ideal angle-steering patterns $g_{\Theta_{\rx}}(\phi|\hat{\phi}_{\text{c}})$ and $g_{\Theta_{\tx}}(\theta|\hat{\theta}_{\text{c}})$ pointing at $\hat{\phi}_{\text{c}}$, $\hat{\theta}_{\text{c}}$ with beam-width $\Theta_{\rx}$ and $\Theta_{\tx}$, respectively. The ideal pattern is defined by rectangular gain envelope and unit power in angular domain, i.e.,
\begin{align}
g_{\Theta_{\rx}}\left(\phi|\hat{\phi}_{\text{c}}\right) = 
\begin{cases}
\sqrt{\pi/\Theta_\rx}, & |\phi-\hat{\phi}_{\text{c}}| \leq \Theta_{\rx}/2\\
0, & \text{otherwise}
\end{cases}
\label{eq:ideal_pattern}
\end{align}
The average beamforming gain is
\begin{align}
G = \mathbb{E}_{\veta_{\text{r}}}\left\{\left|\sum_{r=1}^{R}\frac{e^{\psi_{r}}}{\sqrt{R}}g_{\Theta_{\rx}}(\phi_{\text{c}} + \Delta\phi_r|\hat{\phi}_{\text{c}})g_{\Theta_{\rx}}(\theta_{\text{c}} + \Delta\theta_r|\hat{\theta}_{\text{c}})\right|^2\right\}
\label{eq:beamforming_gain}
\end{align}

In this work, our goal is to:
\begin{itemize}
\item Determine the error bound in tracking propagation directions, i.e., $\text{var}(\theta_{\epsilon})$ and $\text{var}(\phi_{\epsilon})$, where $\theta_{\epsilon}= \theta_{\text{c}} -\hat{\theta}_{\text{c}}$ and $\theta_{\epsilon}= \phi_{\text{c}} -\hat{\phi}_{\text{c}}$ with respect to AS.
\item Determine the average gain $G$ from data phase beamforming and its relationship with angle estimation error $\phi_{\epsilon}, \theta_{\epsilon}$, beam-widths and AS. 
\item Compare the gain obtained from two practical angle tracking algorithms (ML channel tracking algorithm and sector tracking) with respect to different beam-widths.
\end{itemize}

%
%
\section{Tracking Performance Analysis}
\label{sec:CRLB}
In this section, we provide Cramer-Rao lower bound (CRLB) of the angle estimators $\hat{\phi}_c$ and $\hat{\theta}_c$, which are lower bounds of variance of $\phi_{\epsilon}$ and $\theta_{\epsilon}$ with unbiased estimators. We also provide theoretical analysis of beamforming gain $G$.
\subsection{CRLB of angle tracking accuracy}

We define $\veta = [\veta_{\text{c}}^{\transpose},\veta^{\transpose}_{\text{r}}]^{\transpose}$ that contains both cluster-specific and ray-specific parameters. The error bound of $\veta$ is
\begin{align}
\mathbb{E}_{\mathbf{y},\veta}\left\{(\veta-\hat{\veta})(\veta-\hat{\veta})^{\transpose}\right\}\succcurlyeq \mathbf{J}_{\veta}^{-1},
\end{align}
where $\mathbf{J}_{\veta}$ is the Fisher information matrix (FIM) of parameters, and it is defined as
\begin{align}
\mathbf{J}_{\veta} = -\mathbb{E}_{\mathbf{y},\veta}\left\{\frac{\partial^2}{\partial \veta \partial \veta^{\transpose}}\ln f(\mathbf{y},\veta) \right\}.
\end{align}

The likelihood function is $f(\mathbf{y},\veta) = f(\mathbf{y}|\veta) f_{\text{p}} (\veta)$ which contains a-priori probability of $\veta$, i.e., $f_{\text{p}}(\veta)$, and conditional probability $f(\mathbf{y}|\veta)$ from thermal noise. The FIM is the sum of the FIM from observation $\mathbf{J}_\text{w}$ and the FIM from a-priori of parameters $\mathbf{J}_\text{p}$ \cite{5571900}
\begin{align}
\mathbf{J}_{\veta} = \mathbf{J}_\text{w}+\mathbf{J}_{\text{p}}.
\end{align}
The elements of $\mathbf{J}_\text{w}$ are defined as
\begin{align}
\{\mathbf{J}_{\text{w}}\}_{m,n} = &\mathbb{E}_{\mathbf{y}|\veta} \left\{\Re \left[\frac{\partial (-\ln f(\mathbf{y}|{\veta}))}{\partial \veta_m}\frac{\partial (-\ln f^{\hermitian}(\mathbf{y}|{\veta}))}{\partial \veta_n}\right]\right\}\nonumber\\
= & \begin{bmatrix}
\mathbf{J}_{\text{c}} & \mathbf{J}_{\text{c,r}} \\
\mathbf{J}^{\transpose}_{\text{c,r}} & \mathbf{J}_{\text{r}}
\end{bmatrix},
\end{align}
where $\veta_m$ and $\veta_n$ are the $m^{\text{th}}$ and $n^{\text{th}}$ element of vector $\veta$. The exact expression is shown in the Appendix~\ref{appendix:FIM}. The elements of $\mathbf{J}_{\text{p}}$ are defined as
\begin{align}
\{\mathbf{J}_{\text{p}}\}_{m,n} = &\mathbb{E}_{\veta} \left\{\Re \left[\frac{\partial (-\ln f_{\text{p}}({\veta}))}{\partial \veta_m}\frac{\partial (-\ln f_{\text{p}}^{\hermitian}({\veta}))}{\partial \veta_n}\right]\right\}.
\end{align}
Due to independent assumption among all parameters, $\mathbf{J}_{\text{p}}$ is a diagonal matrix. By using the knowledge of a-priori probability of AO\footnote{Although AO is typically modeled as Laplacian distribution, we use Gaussian distribution for mathematical convenience.}, $f_{\text{p}}(\Delta\phi_r) = p_{\text{AO}}(\Delta\phi_r)$ and $f_{\text{p}}(\Delta\theta_r) = p_{\text{AO}}(\Delta\theta_r)$, the matrix $\mathbf{J}_{\text{p}}$ becomes
\begin{align}
\mathbf{J}_{\text{p}} = \sigma^{-2}_{\text{RAS}}
\begin{bmatrix}
\mathbf{0} & \mathbf{0}\\
\mathbf{0} & \mathbf{E}_1
\end{bmatrix}+
\sigma^{-2}_{\text{TAS}}
\begin{bmatrix}
\mathbf{0} & \mathbf{0}\\
\mathbf{0} & \mathbf{E}_2
\end{bmatrix}
\end{align}
where $\mathbf{E}_1 = \mathbf{I}_R \otimes \text{diag}([1,0,0]^{\transpose})$ and $\mathbf{E}_2 = \mathbf{I}_R \otimes \text{diag}([0,1,0]^\transpose)$. The operator $\otimes$ is the Kronecker product.

Note that we use the definition of the equivalent FIM (EFIM) \cite{5571900} in order to express the FIM as a block matrix of $\veta_{\text{c}}$ and $\veta{\text{r}}$.
It allows us to determine the CRLB of cluster-specific parameter, and the CRLBs of AoA and AoD are
\begin{align}
\text{var}(\hat{\phi}_\text{c}) \geq \{\mathbf{J}^{-1}_{\veta_{\text{c}}}\}_{1,1},\text{var}(\hat{\theta}_\text{c}) \geq \{\mathbf{J}^{-1}_{\veta_{\text{c}}}\}_{2,2},
\label{eq:CRLB}
\end{align}
where $\mathbf{J}^{-1}_{\veta_{\text{c}}}
= \left(\mathbf{J}_{\text{c}} -\mathbf{J}^{\transpose}_{\text{c,r}}\left(\mathbf{J}_{\text{r}}+\sigma^{-2}_{\text{RAS}}\mathbf{E}_1+\sigma^{-2}_{\text{TAS}}\mathbf{E}_2\right)^{-1}\mathbf{J}_{\text{c,r}}\right)^{-1}.$

\subsection{Data phase beamforming gain with angle steering}
In this subsection, we study the impact of angle tracking errors $\phi_{\epsilon}$, $\theta_{\epsilon}$ and beam-widths $\Theta_{\rx}, \Theta_{\tx}$ on beamforming gain in the data phase.

\textit{Proposition 1:} The average data phase gain is 
\begin{align}
G =  G_{\tx}(\theta_{\epsilon})
G_{\rx}(\phi_{\epsilon})
\label{eq:proposition_gain}
\end{align}
where the average receiver and transmitter gain $G_{\rx}$ and $G_{\tx}$ are convolutions between the beam pattern and a PDF of corresponding angular spread $p_{\text{AS}}(x)$, where $x \in \{\phi, \theta\}$
\begin{align}
G_{\rx}(\phi) = & g^2_{\Theta_{\rx}}\left(\phi|0\right)*p_{\text{AS}}(\phi)\nonumber\\
G_{\tx}(\theta) =& g^2_{\Theta_{\tx}}\left(\theta|0\right)*p_{\text{AS}}(\theta).
\end{align}

\textit{Proof:} See Appendix~\ref{appendix:gain}.

The above proposition reveals the relationship of beamforming gain with cluster angle estimation error and steering beam-width.
%
%
\section{Practical Angle Tracking Algorithms}
\label{sec:algorithm}
In this section, we study two tracking algorithms from literature. 

\textit{ML Channel Tracking:}
The first group of algorithms which intend to track propagation angle \cite{MRM16_GD_tracking}. The key idea is to refine channel gain $\hat{g}_c$ and angle estimates $\hat{\phi}_c$ and $\hat{\theta}_c$ in each tracking frame. Denote the estimated channel parameter in $n^{\text{th}}$ tracking frame as $\hat{g}_c^{(n)}$, $\hat{\phi}_{\text{c}}^{(n)}$, and $\hat{\theta}_{\text{c}}^{(n)}$. The estimation steps for complex gain $g$ and propagation angles are given by
\begin{align*}
&\hat{g}_c^{(n)} = \text{arg} \min_{g} \left\|\mathbf{y}^{(n)} - g\mathbf{W}^{\text{H}} \mathbf{a}_{\rx}(\hat{\phi}_{\text{c}}^{(n-1)})\mathbf{a}^{\hermitian}_{\tx}(\theta^{(n-1)}))\mathbf{F}\right\|^2\\
&\{\hat{\phi}_{\text{c}}^{(n)}, \hat{\theta}_{\text{c}}^{(n)}\} = \text{arg} \min_{\phi,\theta} \left\|\mathbf{y}^{(n)} - \hat{g}_c^{(n)} \mathbf{W}^{\text{H}} \mathbf{a}_{\rx}(\phi)\mathbf{a}^{\hermitian}_{\tx}(\theta)\mathbf{F}\right\|^2
\end{align*}
assuming initial estimates $\hat{g}_c^{(0)}$, $\hat{\phi}_{\text{c}}^{(0)}$, and $\hat{\theta}_{\text{c}}^{(0)}$ are known. In each tracking frame, the CSI is iteratively refined.

\textit{Sector Tracking:}
In sector tracking, the BS and UE steer sector beams adjacent to the previously used sector beams in order to measure signal strength of transmitter and receiver beam pairs. The tracking algorithm tends to find the best beam pairs that results in the highest SNR and keep on updating the information. Sector size is typically chosen to accommodate beam-width supported by antenna array.

%
\section{Numerical Evaluation}
\label{sec:simulation results}
In this section, we evaluate the tracking performance. In the simulation we use channel model in (\ref{eq:channel_model}), where the centroids of AoA and AoD are uniformly chosen from $[-\pi/2,\pi/2]$. The angular offsets $\Delta\phi_r$ and $\Delta\theta_r$ are zero mean Gaussian random variables. In our simulations, we focus on the AOA tracking and thus set variance of $\Delta\theta_r$ to be $\sigma^2_{\text{TAS}}=0$. The point-to-point SNR is defined in terms of path gain, i.e., $\text{SNR}  = g_{\text{c}}^2/\sigma^2_\text{n}$. The CRLB is computed in a deterministic manner except when a random beamformer is used. For the random beamformer, i.e. when each element in $\mathbf{W}$ and $\mathbf{F}$ is randomly chosen from $\{\pm1\pm1j\}/\sqrt{2N_{\rx}}$ and $\{\pm1\pm1j\}/\sqrt{2N_{\tx}}$, respectively \cite{MRM16_GD_tracking}, the CRLB is obtained by averaging random matrices $\mathbf{W}$ and $\mathbf{F}$ in order to evaluate the average performance. Both BS and UE has 32 antennas in order to flexibly use different beam-width. Steering vectors that adapt beam-width in the data phase follow the codebook from (21) of \cite{6847111}.

\begin{figure}
\begin{center}
\includegraphics[width=0.5\textwidth]{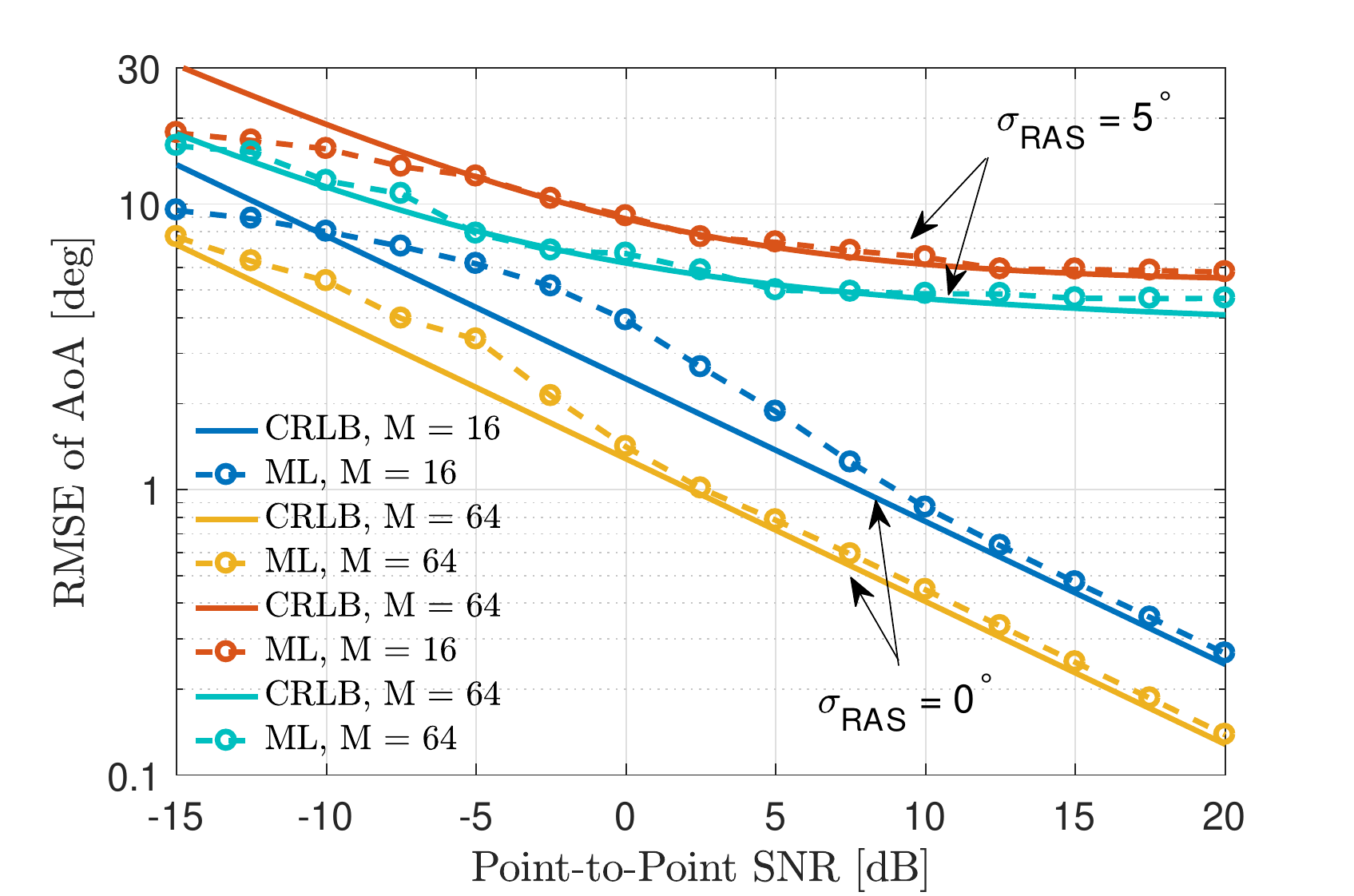}
\end{center}
\vspace{-3mm}
\caption{RMSE of cluster AoA tracking as a function of point-to-point SNR. Different AoA AS $\sigma_{\text{RAS}}$ and training length $M$ are considered.}
\vspace{-5mm}
\label{fig:angle_CRLB}
\end{figure}

In Fig.~\ref{fig:angle_CRLB}, we evaluate the root mean square error (RMSE) of cluster AoA from the ML tracking approach and CRLB from (\ref{eq:CRLB}). Quasi-omni directional (random) training beamformers $\mathbf{W}$ and $\mathbf{F}$ are used for both simulation and computation of the CRLB. Our results show that the ML reaches the CRLB in high SNR regime as expected. Without AS, the cluster AoA estimation accuracy increases with SNR in dB scale, and doubling the number of training slots provides 3dB SNR improvement. When AoA AS is present, the cluster AoA accuracy bound is close to $\sigma_{\text{RAS}}$. Higher SNR and more training slots provide marginal benefits.

\begin{figure}
\begin{center}
\includegraphics[width=0.5\textwidth]{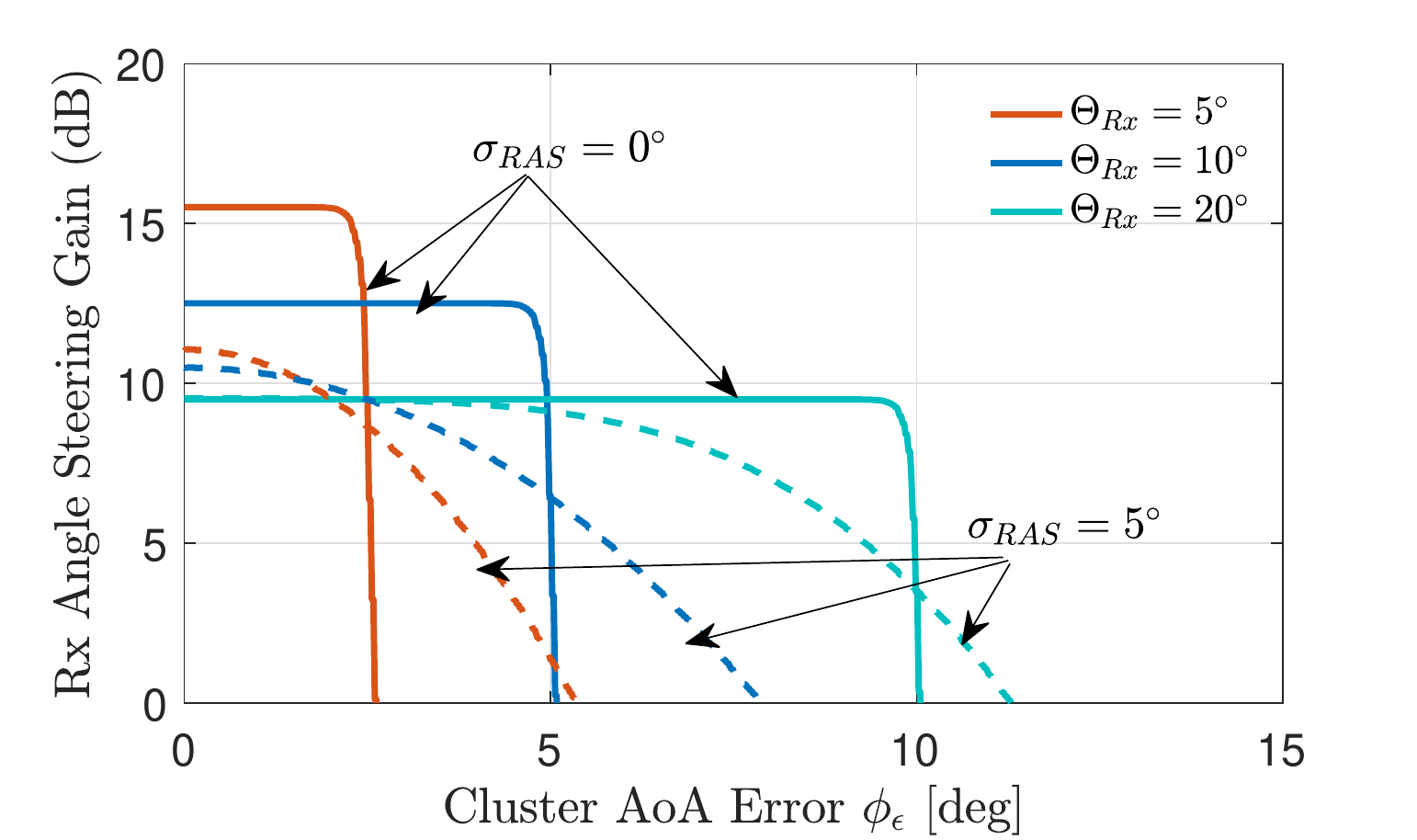}
\end{center}
\vspace{-3mm}
\caption{Average receiver gain as function of beam pointing error $\phi_{\epsilon}$. Different AoA AS $\sigma_{\text{RAS}}$ and beam-widths $\Theta_{\rx}$ are considered.}
\vspace{-4mm}
\label{fig:gain_vs_error}
\end{figure}

In Fig.~\ref{fig:gain_vs_error}, the average receiver beamforming gain $G$ from (\ref{eq:proposition_gain}) is presented as a function of angle steering error $\phi_\epsilon$. When there are no AS and steering errors, narrower beams always provide better beamforming gain. Since the steering error is less troublesome according to results in Fig.~\ref{fig:angle_CRLB}, narrower beams are preferred. However, channel with clustered multipaths does not benefits from it. In dashed curves, angle steering gain does not necessarily increases when beam-width becomes narrower. Additionally, with steering error associated with tracking in clustered multipaths, performance difference between different beams becomes marginal, e.g., from $10^{\circ}$ to $5^{\circ}$.

\begin{figure}[t]
\begin{tabular}{cc}
\subfloat[Simulated AoA tracking performance over 1s duration.]{%
  \includegraphics[clip,width=1\columnwidth]{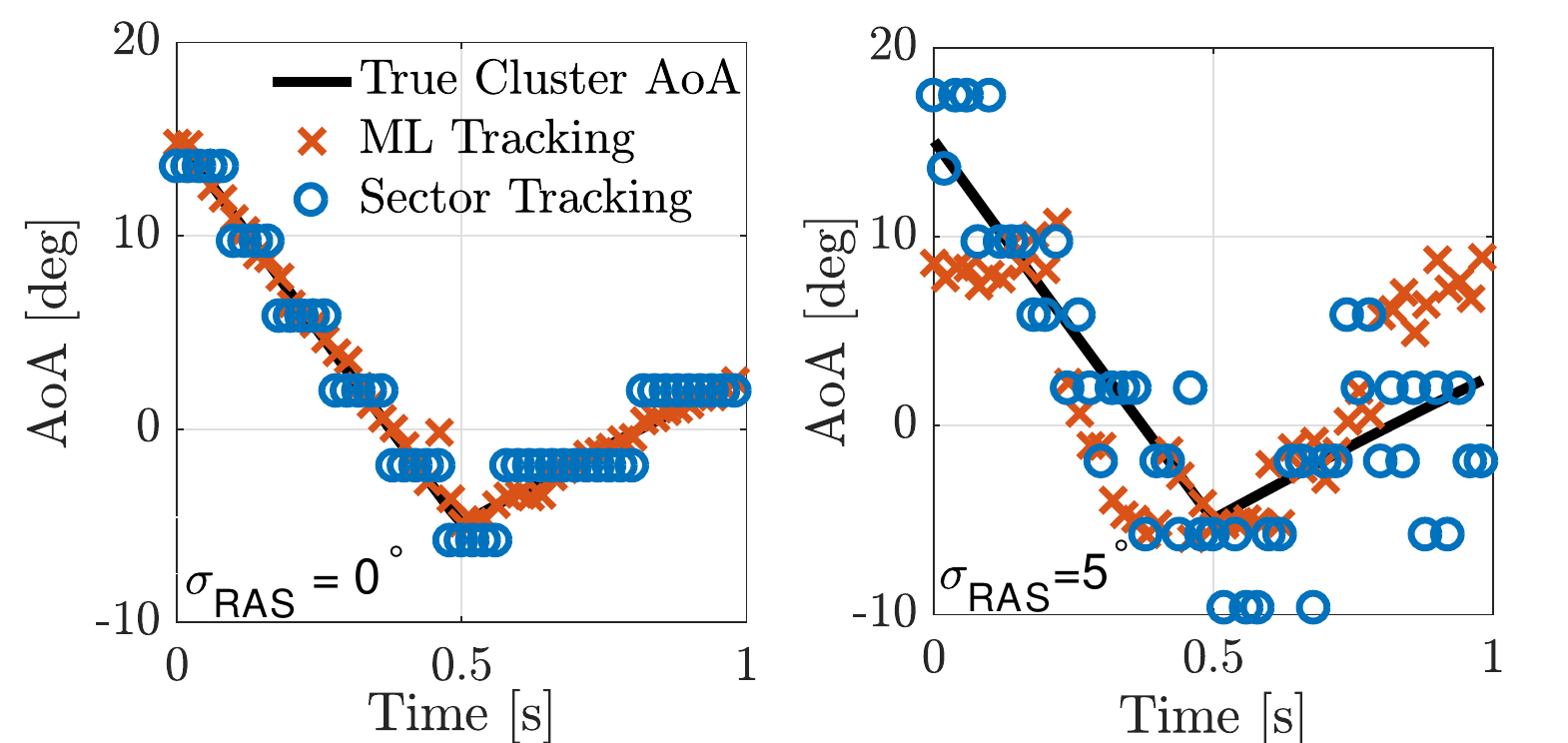}%
}\\
\vspace{-0mm}
\subfloat[Simulated AoA angular power profile of channel over 1s duration.]{%
  \includegraphics[clip,width=1\columnwidth]{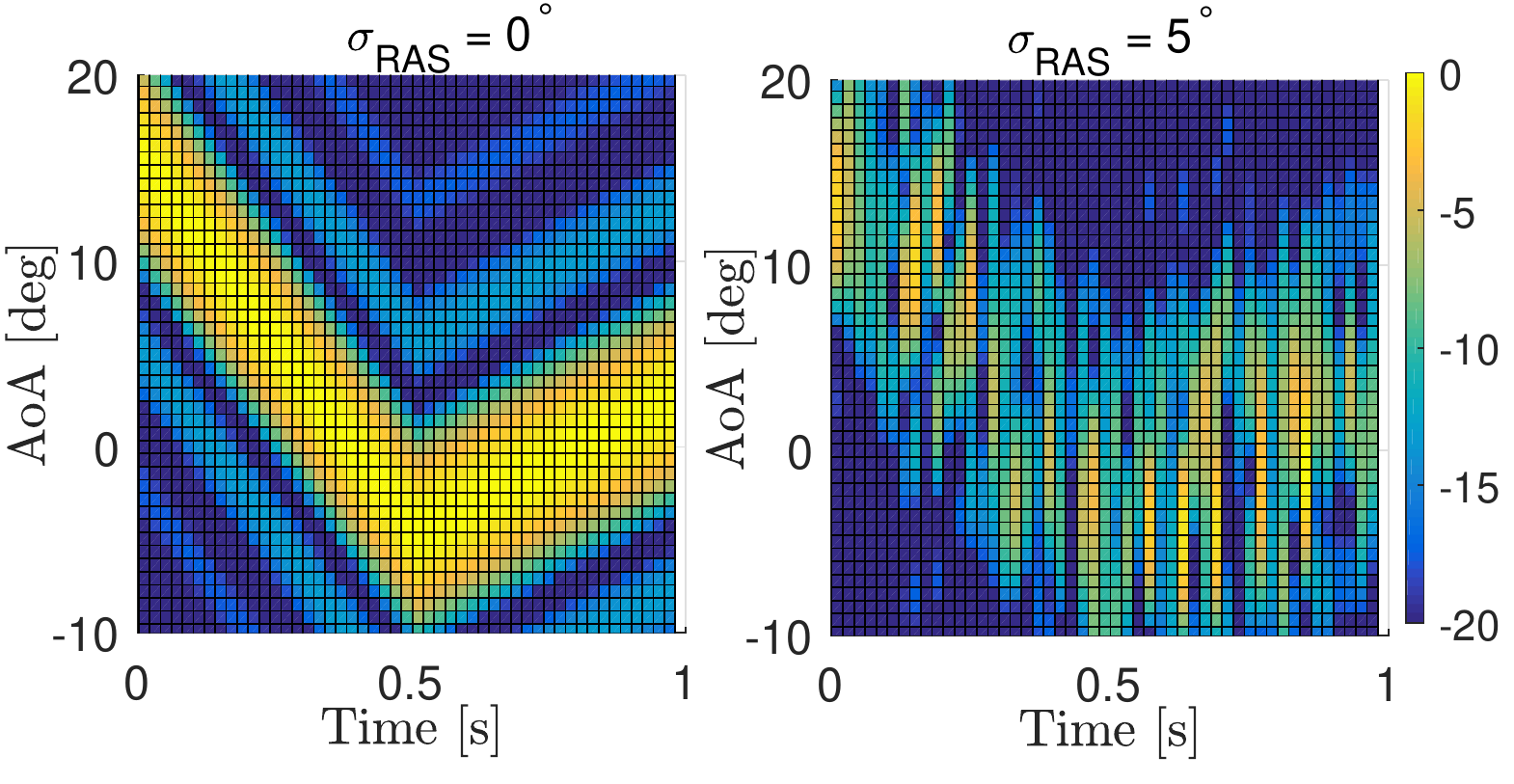}%
}\\
\vspace{-1mm}
\subfloat[CCDF of gain in data phase. Different AoA AS and beam-width are considered.]{%
  \includegraphics[clip,width=1\columnwidth]{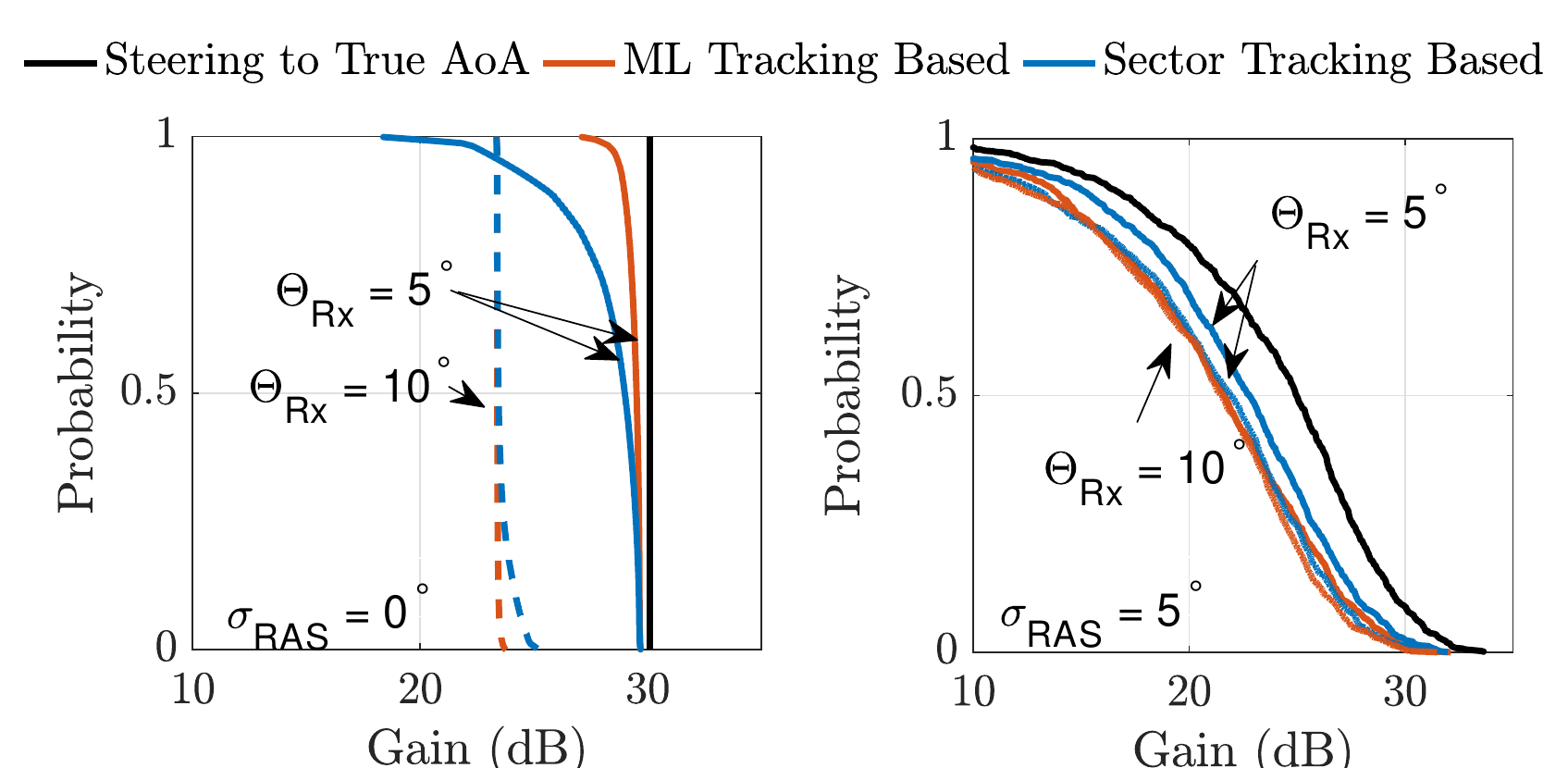}%
}
\end{tabular}
\caption{Simulation example of channel tracking. In the first 0.5s of simulation, UE moves with speed of [10,0]m/s in x/y axis and has -~50deg/s rotational speed. In the second 0.5s, it moves with speed of [0,10]m/s and has 25deg/s rotational speed. Point-to-point SNR is 0dB. }
\vspace{-5mm}
\label{fig:gain_vs_time}
\end{figure}

In Fig.~\ref{fig:gain_vs_time}(a), we evaluate AoA tracking accuracy of both algorithms from Section~\ref{sec:algorithm}. Without AS, the ML tracking outperforms the sector tracking in terms of accuracy since the precision of the latter is limited by the sector size. With AS, tracking accuracy in both algorithms is degraded. The reason is explained in Fig.~\ref{fig:gain_vs_time}(b) which shows angular power profile over time. The angle tracking error in each frame is associated with different realization of intra-cluster angular offsets, and thus it cannot be improved by higher SNR or longer training. In Fig.~\ref{fig:gain_vs_time}(c), we evaluate the complementary cumulative distribution function (CCDF) of achieving a certain gain in the data phase by using these tracking algorithms, and different beam-width in angle steering are considered. For comparison, we also include a benchmark CCDF curve for angle steering in the true cluster AoA. Without AS, which corresponds to LOS, the benchmark has 30dB gain. The ML tracking provides higher gain as compared to the sector tracking due to better accuracy. Steering with $5^\circ$ beam-width has 3dB higher gain over using $10^\circ$. With AS that corresponds to an NLOS path cluster, gain is not fixed due to fading among intra-cluster rays. Both tracking algorithms have worse performance than the benchmark curve due to the error and angular spread. The advantage of ML over sector tracking becomes marginal and steering narrow beams does not improve the gain. Due to the relationship between the number of antenna elements and the narrowest possible beam, designer should take angular spread into account when the number of UE antenna elements is considered.


%
%
\section{Conclusion}
\label{sec:Conclusion}
In this work, we study the performance bound of narrowband channel tracking techniques under clustered sparse channel. We provide the Cramer-Rao lower bound of angle tracking error. Our study reveals that the non-clustered sparse model degrades the angle tracking accuracy. With clustered multipaths, the angle tracking accuracy is bounded by cluster angular spread of propagation path and cannot be further improved by increasing the signal strength or the training duration. We also analyzed the beamforming gain using angle-steering-based beamformer and showed that under angular spread narrow beams do not necessarily provide better gain.


\appendix
\subsection{FIM expression of channel parameters}
\label{appendix:FIM}
Likelihood function of $\mathbf{y}$ given $\veta$ is multivariate Gaussian PDF. We define the following terms $\mathbf{d}_{\rx}(\phi) = \frac{\partial \mathbf{a}_{\rx}(\phi)}{\partial \phi_\text{c}}$ and $\mathbf{d}_{\tx}(\theta) = \frac{\partial \mathbf{a}_{\tx}(\theta)}{\partial \theta}$.
We further define $\tilde{\mathbf{a}}_{\rx}(\phi) = \mathbf{W}^{\hermitian}\mathbf{a}_{\rx}\left(\phi\right)$, $\tilde{\mathbf{d}}_{\rx}(\phi) = \mathbf{W}^{\hermitian}\mathbf{d}_{\rx}\left(\phi\right)$, $\tilde{\mathbf{a}}_{\tx}(\theta) = \mathbf{F}^{\hermitian}\mathbf{a}_{\tx}\left(\theta\right)$ and $\tilde{\mathbf{d}}_{\tx}(\theta) = \mathbf{F}^{\hermitian}\mathbf{d}_{\tx}\left(\theta\right)$.
We also use the following definitions for clarity 
\begin{align}
\begin{split}
\mathbf{v}_{\text{aa},r} = \text{diag}\left[e^{j\psi_r}\tilde{\mathbf{a}}_{\rx}\left(\phi_{\text{c}}+\Delta\phi_r\right)\tilde{\mathbf{a}}^{\hermitian}_{\tx}\left(\theta_{\text{c}}+\Delta\theta_r\right)\right],\\
\mathbf{v}_{\text{da},r} = \text{diag}\left[e^{j\psi_r}\tilde{\mathbf{d}}_{\rx}\left(\phi_{\text{c}}+\Delta\phi_r\right)\tilde{\mathbf{a}}^{\hermitian}_{\tx}\left(\theta_{\text{c}}+\Delta\theta_r\right)\right],\\
\mathbf{v}_{\text{ad},r} = \text{diag}\left[e^{j\psi_r}\tilde{\mathbf{a}}_{\rx}\left(\phi_{\text{c}}+\Delta\phi_r\right)\tilde{\mathbf{d}}^{\hermitian}_{\tx}\left(\theta_{\text{c}}+\Delta\theta_r\right)\right],\\
\mathbf{v}_{\text{dd},r} = \text{diag}\left[e^{j\psi_r}\tilde{\mathbf{d}}_{\rx}\left(\phi_{\text{c}}+\Delta\phi_r\right)\tilde{\mathbf{d}}^{\hermitian}_{\tx}\left(\theta_{\text{c}}+\Delta\theta_r\right)\right].
\end{split}
\label{eq:appendix_def2}
\end{align}
Besides, we define $\mathbf{v}_{\text{a,a}} = \sum_{r=1}^{R}\mathbf{v}_{\text{aa},r}$,$\mathbf{v}_{\text{d,a}} = \sum_{r=1}^{R}\mathbf{v}_{\text{da},r}$ and $\mathbf{v}_{\text{a,d}} =\sum_{r=1}^{R}\mathbf{v}_{\text{ad},r}$. We then define the following notation where $x$ and $y$ can be arbitrary parameters in $\veta$
\begin{align}
&J_{x,y}=J_{y,x}=-\frac{\sigma^2_{\text{n}}}{2}\mathbb{E}_{\mathbf{y},\veta}\left[\frac{\partial^2\ln f(\mathbf{y}|\veta)}{\partial x \partial y}\right],
\end{align}

The elements in FIM $\mathbf{J}_{\text{c}} \in \mathbb{R}^{3\times 3}$ are
\begin{align}
\mathbf{J}_{\text{c}}= \frac{2}{\sigma^2_{\text{n}}}
\begin{bmatrix}
J_{\phi_{\text{c}},\phi_{\text{c}}} & J_{\phi_{\text{c}},\theta_{\text{c}}} & J_{\phi_{\text{c}},g_{\text{c}}} \\
J_{\theta_{\text{c}},\phi_{\text{c}}} & J_{\theta_{\text{c}},\theta_{\text{c}}} & J_{\theta_{\text{c}},g_{\text{c}}} \\
J_{g_{\text{c}},\phi_{\text{c}}} & J_{g_{\text{c}},\theta_{\text{c}}} & J_{g_{\text{c}},g_{\text{c}}} \\
\end{bmatrix}
\end{align}
where $J_{\phi_{\text{c}},\phi_{\text{c}}}=\Re\{\mathbf{v}^{\hermitian}_{\text{d,a}}\mathbf{v}_{\text{d,a}}\},$ $J_{\theta_{\text{c}},\theta_{\text{c}}}=\Re\{\mathbf{v}^{\hermitian}_{\text{a,d}}\mathbf{v}_{\text{a,d}}\}$ $J_{\phi,\theta}=\Re\{\mathbf{v}^{\hermitian}_{\text{d,a}}\mathbf{v}_{\text{a,d}}\},$ $J_{g,\phi}=\Re\{\mathbf{v}^{\hermitian}_{\text{a,a}}\mathbf{v}_{\text{d,a}}\},$ $J_{g,\theta}=\Re\{\mathbf{v}^{\hermitian}_{\text{a,a}}\mathbf{v}_{\text{a,d}}\},$ $J_{\alpha,\alpha}=\Re\{\mathbf{v}^{\hermitian}_{\text{a,a}}\mathbf{v}_{\text{a,a}}\}.$

Next, we evaluate elements in FIM $\mathbf{J}_{\text{r}} \in \mathbb{R}^{3R\times 3R}$ by the following block division
\begin{align}
\mathbf{J}_{\text{r}}=\frac{2}{\sigma^2_{\text{n}}}
\begin{bmatrix}
\mathbf{J}_{\text{r}}^{(1,1)}  &  \cdots & \mathbf{J}_{\text{r}}^{(1,R)} \\
\vdots & \ddots & \vdots \\
\mathbf{J}_{\text{r}}^{(R,R)}  &  \cdots & \mathbf{J}_{\text{r}}^{(R,R)} \\
\end{bmatrix}
\end{align}
where
\begin{align}
\mathbf{J}^{(m,n)}_{\text{r}}=
\begin{bmatrix}
J_{\phi_m,\phi_n}  & J_{\phi_m,\theta_n} & J_{\phi_m,\psi_n} \\
J_{\theta_m,\phi_n}  & J_{\theta_m,\theta_n} & J_{\theta_m,\psi_n} \\
J_{\psi_m,\phi_n} & J_{\psi_m,\theta_n} & J_{\psi_m,\psi_n} 
\end{bmatrix}
\label{eq:J_r}
\end{align}
The specific expressions in (\ref{eq:J_r}) are $J_{\phi_m,\phi_n}=\Re\{\mathbf{v}^{\hermitian}_{\text{da},m}\mathbf{v}_{\text{da},n}\},$ $J_{\phi_m,\theta_n}=\Re\{\mathbf{v}^{\hermitian}_{\text{da},m}\mathbf{v}_{\text{ad},n}\}$ $J_{\theta_m,\phi_n}=\Re\{\mathbf{v}^{\hermitian}_{\text{ad},m}\mathbf{v}_{\text{da},n}\},$ and $J_{\theta_m,\theta_n}=\Re\{\mathbf{v}^{\hermitian}_{\text{ad},m}\mathbf{v}_{\text{ad},n}\}.$

The elements in FIM $\mathbf{J}_{\text{c,r}}$ are shown in block division
\begin{align}
\mathbf{J}_{\text{c,r}}=\frac{2}{\sigma_\text{\text{n}}^2}
\left[\mathbf{J}_{\text{c,r}}^{(1)},\cdots,\mathbf{J}_{\text{c,r}}^{(R)}\right]^{\transpose},
\end{align}
where
\begin{align}
\mathbf{J}_{\text{c,r}}^{(r)}=
\begin{bmatrix}
J_{\phi_r,\phi_{\text{c}}} & J_{\phi_r,\theta_{\text{c}}} & J_{\phi_r,g_{\text{c}}}  \\
J_{\theta_r,\phi_{\text{c}}} & J_{\theta_r,\theta_{\text{c}}} & J_{\theta_r,g_{\text{c}}} \\
J_{\psi_r,\phi_{\text{c}}} & J_{\psi_r,\theta_{\text{c}}} & J_{\psi_r,g_{\text{c}}}
\end{bmatrix}.
\label{eq:appendix_Jd0}
\end{align}
The specific expressions in (\ref{eq:appendix_Jd0}) are $J_{\phi_r,\phi_{\text{c}}}=\Re\{\mathbf{v}^{\hermitian}_{\text{da},r}\mathbf{v}_{\text{d,a}}\},$ $J_{\phi_r,\theta_{\text{c}}}=\Re\{\mathbf{v}^{\hermitian}_{\text{da},r}\mathbf{v}_{\text{a,d}}\},$ $J_{\phi_r,g_{\text{c}}}=\Re\{\mathbf{v}^{\hermitian}_{\text{da},r}\mathbf{v}_{\text{a,a}}\},$ $J_{\theta_r,\phi_{\text{c}}}=\Re\{\mathbf{v}^{\hermitian}_{\text{ad},r}\mathbf{v}_{\text{da}}\},$ $J_{\theta_r,\theta_{\text{c}}}=\Re\{\mathbf{v}^{\hermitian}_{\text{ad},r}\mathbf{v}_{\text{ad}}\},$ $J_{\theta_r,\phi_{\text{c}}}=\Re\{\mathbf{v}^{\hermitian}_{\text{ad},r}\mathbf{v}_{\text{aa}}\},$ and $J_{\theta_r,\theta_{\text{c}}}=\Re\{\mathbf{v}^{\hermitian}_{\text{ad},r}\mathbf{v}_{\text{aa}}\}.$

\subsection{Average beamforming gain under angular spread}
\label{appendix:gain}
The average gain in (\ref{eq:beamforming_gain}) can be expressed as:
\begin{align*}
&G = \mathbb{E}_{\veta_{\text{r}}}\bigg\{\sum_{r_1=1}^{R}\sum_{r2=1}^{R}\frac{e^{j(\psi_{r_1}-\psi_{r_2})}}{R}g_{\Theta_{\rx}}(\phi_{\text{c}} + \Delta\phi_{r_1}|\hat{\phi}_{\text{c}})\nonumber\\
&g_{\Theta_{\tx}}(\theta_{\text{c}} + \Delta\theta_{r_1}|\hat{\theta}_{\text{c}})g_{\Theta_{\rx}}(\phi_{\text{c}} + \Delta\phi_{r_2}|\hat{\phi}_{\text{c}})g_{\Theta_{\tx}}(\theta_{\text{c}} + \Delta\theta_{r_2}|\hat{\theta}_{\text{c}})\bigg\}
\end{align*}

The expectation over $\psi_r$ reduces the expression to 
\begin{align*}
G = \underbrace{\mathbb{E}_{\Delta\phi_r}\bigg\{g^2_{\Theta_{\rx}}(\phi_{\text{c}} + \Delta\phi_{r}|\hat{\phi}_{\text{c}})\bigg\}}_{G_{\rx}(\theta_{\epsilon})}
\underbrace{\mathbb{E}_{\Delta\theta_r}\bigg\{g^2_{\Theta_{\tx}}(\theta_{\text{c}} + \Delta\theta_{r}|\hat{\theta}_{\text{c}})\bigg\}}_{G_{\tx}(\phi_{\epsilon})}
\end{align*}
since expectations are zero except for $r_1=r_2$. Using the PDF of AO and definition of ideal beam pattern (\ref{eq:ideal_pattern}), the Rx gain $G_{\rx} (\phi_{\epsilon})$ is 
\begin{align*}
G_{\rx}(\phi_{\epsilon}) 
=  \frac{\pi}{\Theta_{\rx}}\int_{-\phi_{\epsilon}-\frac{\Theta_{\rx}}{2}}^{-\phi_{\epsilon}+\frac{\Theta_{\rx}}{2}} p_{\text{AO}}(\phi)d\phi
\end{align*}
which is equivalent to $[g^2_{\Theta_{\rx}}\left(\phi|0\right)*p_{\text{AO}}(\phi)]|_{\phi=\phi_{\epsilon}}$. The Tx gain is derived similarly, and then expression (\ref{eq:proposition_gain}) is obtained.
%
%



%
\bibliographystyle{IEEEtran}
\bibliography{IEEEabrv,references}

\end{document}